\documentclass[twocolumn,english,showpacs,preprintnumbers,amsmath,amssymb,floatfix]{revtex4}
\usepackage[T1]{fontenc}
\usepackage{color}
\usepackage{array}
\usepackage{amstext}
\usepackage{graphicx}
\usepackage{esint}
\usepackage[colorlinks = true,
linkcolor = magenta,
urlcolor  = blue,
citecolor = red,
anchorcolor = blue]{hyperref}

\makeatletter

\@ifundefined{textcolor}{}
{%
	\definecolor{BLACK}{gray}{0}
	\definecolor{WHITE}{gray}{1}
	\definecolor{RED}{rgb}{1,0,0}
	\definecolor{GREEN}{rgb}{0,1,0}
	\definecolor{BLUE}{rgb}{0,0,1}
	\definecolor{CYAN}{cmyk}{1,0,0,0}
	\definecolor{MAGENTA}{cmyk}{0,1,0,0}
	\definecolor{YELLOW}{cmyk}{0,0,1,0}
}

\@ifundefined{definecolor}
{\usepackage{color}}{}
\@ifundefined{definecolor}
{\usepackage{color}}{}
\makeatother

\makeatother

\usepackage{babel}
\begin{document}
	
	\title{QCD analysis of light charged Higgs production through polarized top quark decay in two various frames}

	\author{S. Mohammad Moosavi Nejad$^{a,b}$}
	\email{mmoosavi@yazd.ac.ir}
	
	\author{S. Abbaspour$^a$}

	\affiliation{$^{(a)}$Faculty of Physics, Yazd University, P.O. Box
		89195-741, Yazd, Iran\\	
		$^{(b)}$School of Particles and Accelerators,
		Institute for Research in Fundamental Sciences (IPM), P.O.Box
		19395-5531, Tehran, Iran}

\date{\today}

\begin{abstract}

Light and heavy charged Higgs bosons are predicted by many models with an extended Higgs sector such as the two-Higgs-doublet model (2HDM).
Searches for the charged Higgs bosons have been done by the ATLAS and the CMS experiments at the Large Hadron Collider (LHC) in proton-proton collision. However, a definitive search is a program that still has to be carried out so this belongs to the LHC experiments. The experimental observation of charged Higgs bosons would indicate physics beyond the Standard Model.
In the present work we study the ${\cal O}(\alpha_s)$ correction to the energy spectrum of the inclusive bottom-flavored mesons $(X_b)$ in polarized top quark decays into a light charged Higgs boson ($m_{H^+}<m_t$) and a massless bottom quark  followed by the hadronization process $b\to X_b$ in the 2HDM, i.e. $t(\uparrow)\to H^+b\to H^+X_b+Jet$. This spin-dependent energy distribution is  studied in two different helicity coordinate systems. This study could be considered as a new channel to search for the charged Higgs bosons. To present our phenomenological predictions, we restrict ourselves to the unexcluded regions of the MSSM $m_{H^+}-\tan\beta$ parameter space  determined by the recent results of  the CMS  and the ATLAS  collaborations.

\end{abstract}

%\pacs{Valid PACS appear here}% PACS, the Physics and Astronomy
                             % Classification Scheme.
%\keywords{Suggested keywords}%Use showkeys class option if keyword
                              %display desired
\maketitle

\section{Introduction}

Charged Higgs bosons are predicted by several non-minimal Higgs scenarios \cite{higg}, such as models including Higgs triplets \cite{Cheng:1980qt} and two-Higgs-doublet models (2HDM) \cite{Lee:1973iz}. In the 2HDM, as a simplest model,  the Higgs sector of the Standard Model (SM) is extended typically by adding an extra  doublet of complex Higgs fields. In this model, after spontaneous symmetry breaking the particle spectrum includes five physical Higgs bosons: light and heavy CP-even Higgs bosons h and H with $m_H>m_h$, a CP-odd Higgs boson A, plus two charged Higgs bosons $H^\pm$ \cite{Djouadi:2005gj}.
The discovery of a charged Higgs boson would clearly indicate  unambiguous evidence for the presence of new physics beyond the SM.\\
The production and decay modes of  charged Higgs bosons depend on their masses, $m_{H^\pm}$.
At hadron colliders, charged Higgs bosons can be produced through several channels.
In a type-II 2HDM, which is the Higgs sector of the Minimal Supersymmetric Standard Model (MSSM) \cite{Inoue:1982pi}, the main production mode at the
Large Hadron Collider (LHC) for light charged Higgs (with $m_{H^\pm}<m_t$) is through the top quark decay  $t\to bH^+$. 
In this case, the light charged Higgses are produced most frequently via $t\bar{t}$ production.
At the LHC, a cross section of $\sigma(pp\to t\bar{t}X)\approx 1$ (nb) is expected at design energy $\sqrt{S}=14$ TeV  \cite{Langenfeld:2009tc}. With the LHC design luminosity of $10^{34} cm^{-2}s^{-1}$ in each of the four experiments, it is expected to produce about  90 million $t\bar t$-pairs per year \cite{Moch:2008qy}.  Thus, the LHC is   a superlative top factory which  lets one to search for the charged Higgs bosons  in the subsequent decay products of the top pairs $t\bar t\rightarrow H^\pm W^\mp b\bar b$  and $t\bar t\rightarrow H^\pm H^\mp b\bar b$ 
when $H^\pm$ decays into $\tau$-lepton and neutrino.
For a review of all available  production modes of light charged Higgs  at the LHC, see also \cite{Aoki:2011wd}.

The combined Large Electron-Positron (LEP) experiments have determined a lower limit for the charged Higgs  mass in a type-II 2HDM with $B(H^+\to \tau\nu)=1$ as $m_{H^+}>94$~GeV \cite{Abbiendi:2013hk}, and the lower limit for any $B(H^+\to\tau\nu)$ as 80~GeV. The experimental results from the Tevatron placed upper limits on $B(t\to H^+ b)$ in the $15-20\%$ range for light charged Higgs bosons. Both the CMS \cite{Chatrchyan:2012vca} and ATLAS \cite{Aad:2012tj,Aad:2012rjx} collaborations  searched for light charged Higgs bosons assuming $B(H^+\to \tau\nu)=1$ and improved the Tevatron limits to the $1-4\%$ range for a mass range $m_{H^+}=90-160$~GeV. 
We will discuss about the recent results on a search for the charged Higgs bosons by the CMS \cite{CMS:2014cdp} and ATLAS \cite{TheATLAScollaboration:2013wia} collaborations when we present our numerical analysis in Sec.~\ref{sec:three}. 

The primary purpose of the present manuscript is the evaluation of the $\alpha_s$-order QCD corrections to the differential decay width ($d\hat\Gamma/dx_i$) of a polarized top quark into a charged Higgs boson and a bottom quark, $t(\uparrow)\to bH^+$, where $x_i$ is the scaled-energy fraction of the b-quark or the gluon emitted at the next-to-leading order (NLO). In the next section, we shall explain that to obtain the energy distribution of hadrons produced through the top decays in the MSSM, one needs these differential decay widths.
The NLO QCD corrected decay distributions, $\Gamma(t \to bH^+)$, were previously computed in \cite{kadeer}  for the polarized top quarks, and in \cite{Ali:2009sm,Czarnecki,Liud,Li:1990cp} for the unpolarized ones. 
In Ref.~\cite{MoosaviNejad:2011yp}, we calculated the unpolarized  differential decay width $d\hat\Gamma(t\to bH^+)/dx_b$ at NLO and showed that our result after integration over $x_b$ ($0\le x_b\le 1$) is in complete agreement with Refs.~\cite{Ali:2009sm,Czarnecki,Liud} and the corrected version of \cite{Li:1990cp}.
In \cite{MoosaviNejad:2016aad}, we studied the ${\cal O}(\alpha_s)$  radiative corrections to the spin-dependent differential decay rate of the process $t(\uparrow)\to bH^+$ in a special helicity coordinate system with the event plane defined in the $(x, z)$ plane and the $z$-axis along the Higgs boson  three-momentum (in the following called system 1). In this frame, the top quark polarization vector was measured with respect to the direction of the Higgs 3-momentum. 
We checked that our result was in complete agreement with the result presented in \cite{kadeer} after integration over $x_b$ ($0\le x_b\le 1$).\\
Generally, to define the planes one needs to measure the momentum
directions of the momenta $\vec{p}_b$ and $\vec{p}_{H^+}$ and the polarization direction of the top quark, where the measurement of the momentum direction of $\vec{p}_b$ requires the use of a jet finding algorithm, whereas the polarization direction of the top quark must be obtained from the theoretical input. For example, in  $e^+e^-$  interactions the polarization degree of the top quark can be tuned with the help of polarized beams.

In the present work, we analyze the angular distribution of differential width of the process $t(\uparrow)\to bH^+$ in a different helicity coordinate system where, as before, the event plane is the $(x, z)$ plane but with the $z$-axis along the bottom quark (in the following called system 2). In this system, the polarization direction of the top quark is evaluated with respect  to the b-quark three-momentum ($z$-axis). This result is completely new. We also calculate the decay width $\Gamma(t(\uparrow)\to bH^+)$ in this new frame by integrating $d\Gamma/dx_b$ over $0\leq x_b\leq 1$ and compare it with the previous result from \cite{kadeer}.

On the other hand,  bottom quarks produced through the top decays hadronize ($b\to X_b$) before they decay, therefore, each b-jet $X_b$ contains a bottom-flavored hadron which most of the times is a B-meson. At the LHC,  of particular interest is the distribution in the scaled-energy of B-mesons ($x_B$) produced through $t(\uparrow)\to BH^++X$ in the top quark rest frame. The study of these energy distributions in the polarized and unpolarized top decays could be proposed as a new channel to search for the charged Higgs bosons at the LHC.
In \cite{MoosaviNejad:2011yp}, we studied  the energy spectrum of the bottom-flavored mesons in unpolarized top quark decays into a charged Higgs boson and a bottom quark at NLO in the 2HDM. In \cite{MoosaviNejad:2016aad} we studied the spin-dependent energy distribution of B-mesons produced through the polarized top  decays  at NLO in the helicity coordinate system~1. Here, our specific purpose is to study this angular correlation in a different helicity frame  (system~2).
Through this paper, we present our predictions for the B-meson energy  spectrum in the polarized and unpolarized top decays and shall compare
the polarized results in both helicity systems 1 and 2.

In the SM, due to the element $|V_{tb}|\approx 1$ of the Cabibbo-Kobayashi-Maskawa (CKM) \cite{Cabibbo:1963yz}
quark mixing matrix, the top quark  decays dominantly through the two-body mode $t\rightarrow bW^+$. In  \cite{Kniehl:2012mn,Nejad:2013fba,Nejad:2014sla,Nejad:2015pca,Nejad:2016epx}, we investigated the  energy distribution of B-mesons produced in polarized and unpolarized top quark decays in the SM. In each top decay (polarized or unpolarized), to obtain the total distribution of the B-hadron energy   two contributions due to the decay modes $t\rightarrow bH^+$ (in the 2HDM) and $t\rightarrow bW^+$ (in the SM) should be summed up.  Although,
the SM contribution  is normally larger than the one coming from 2HDM \cite{MoosaviNejad:2011yp}.

Finally, We mention that highly polarized top quarks will become available at hadron colliders
through single top production processes, which occur at the $33\%$ level of the $t\bar{t}$ pair production rate \cite{Mahlon:1996pn},
and in top quark pairs produced in future linear $e^+e^-$-colliders \cite{Kuhn:1983ix}.

This paper is organized as follows.
In Sec.~\ref{sec:one}, we study the inclusive production of a meson from polarized top quark considering the factorization theorem and DGLAP equations. In Sec.~\ref{sec:two}, we present our analytical results of the ${\cal O}(\alpha_S)$ QCD corrections to the tree-level rate of $t(\uparrow)\rightarrow bH^+$.
In Sec.~\ref{sec:three}, we present our numerical analysis of inclusive production of a meson from polarized top quark decay considering two different helicity coordinate systems.
In Sec.~\ref{sec:four},  our conclusions are summarized.

\boldmath
\section{Formalism}
\label{sec:one}
\unboldmath

In the proposed way to search for the light charged Higgs bosons, we study the inclusive production of a bottom-flavored meson ($B$) from polarized top quark decay in the following process
\begin{eqnarray}\label{pros}
t(\uparrow)\to bH^+ (g)\rightarrow H^+B+X,
\end{eqnarray}
where $X$ stands for the unobserved final states and the gluon  contributes to the real radiation at NLO. Both the b-quark and the gluon may hadronize into the $B$-meson.

If we label the four-momenta of  top quark, $b$-quark, gluon and $B$-meson by $p_t, p_b, p_g$ and $p_B$, respectively, then  in the top quark rest frame the b-quark, gluon, and B-meson take energies $E_i=p_t\cdot p_i/m_t  (i=b,g,B)$, where $m_B\le E_B\le (m_t^2+m_B^2-m^2_{H^+})/(2m_t) $,  $m_b\le E_b\le (m_t^2+m_b^2-m^2_{H^+})/(2m_t) $ and  $0\le E_g\le (m_t^2-(m_b+m_{H^+})^2)/(2m_t) $.
Following Ref.~\cite{Kniehl:2012mn}, it is convenient to introduce the scaled energy fractions  $x_i=E_i/E_b^{max}=2 E_i/(m_t(1+R-y))$ ($i=b, g, B$) where the scaled  masses $y$ and $R$ are  defined as $y=m_{H^+}^2/m_t^2$ and $R=m_{b}^2/m_t^2$.
By neglecting the b-quark mass $m_b$, one has $x_i=2E_i/(m_t(1-y))$ so that $0\leq (x_b,x_g) \leq 1$.

\begin{figure}
	\begin{center}
		\includegraphics[width=0.45\linewidth,bb=130 10 340 160]{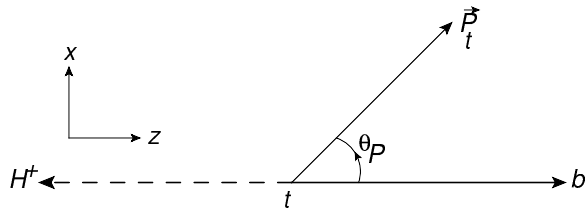}
		\caption{\label{lo}%
			Polar angle $\theta_P$ defined in the second helicity coordinate system (system 2).  $\vec{P}_t$ stands for the top  polarization vector in the top rest frame.}
	\end{center}
\end{figure}

In the first step, we analyze the parton-level sector of the decay process (\ref{pros}) in the rest frame of a top quark.
The angular distribution of the  differential decay width $d\hat\Gamma/dx_i (i=b,g)$  of a polarized top quark is given by the following simple expression
to clarify the correlation between the polarization of the top quark and its decay products
\begin{eqnarray}\label{widthdefine}
\frac{d^2\hat\Gamma(t(\uparrow)\to bH^+ (g))}{dx_i d\cos\theta_P}=\frac{1}{2}\bigg\{\frac{d\hat\Gamma^{unpol}}{dx_i}\pm P\frac{d\hat\Gamma^{pol}}{dx_i}\cos\theta_P\bigg\},
\end{eqnarray}
where $P$ is the polarization degree of the top quark with $0\leq P\leq 1$ so that  $P=1$ corresponds to $100\%$ top quark polarization and
$P=0$ corresponds to an unpolarized top quark.
In Eq.~(\ref{widthdefine})  $d\hat\Gamma^{unpol}/dx_i$ stands for the  unpolarized differential rate, which is extensively calculated in \cite{MoosaviNejad:2011yp} up to NLO, and $d\hat\Gamma^{pol}/dx_i$ refers to the polarized one. The analytical expression for the differential partial width $d\hat\Gamma^{pol}/dx_i$ depends on the selected helicity coordinate system.  
In the rest frame of a top quark decaying into a b-quark, a Higgs boson and a gluon, the final state particles define an event plane. Relative to this  plane, we can define the polarization direction of the polarized top quark. For the decay process  (\ref{pros}), there are two various choices of possible coordinate systems relative to the event plane where one differentiates between frames according to the orientation of the $z$-axis.\\
In \cite{MoosaviNejad:2016aad}, we calculated the angular distribution of the partial decay width $d\hat\Gamma/dx_i$ in a specific frame (system 1) where the three-momentum of the charged Higgs boson ($\vec{P}_{H^+}$) pointed in the direction of the positive $z$-axis and the polar angle $\theta_P$ was defined as the angle between the polarization vector $\vec{P}_t$ of the top quark and the positive $z$-axis. The sign $'+'$ in (\ref{widthdefine}) stands for this system.\\
Here, we consider a different helicity coordinate system (system 2) where the three-momentum of the bottom quark points in the direction of the positive $z$-axis (see Fig.~\ref{lo}).  In (\ref{widthdefine}), the sign $'-'$ stands for the system~2. The  technical detail of our calculation will be presented in the next section. We will show that the results depend on the selected helicity system.

Having the parton-level differential decay rates $d\hat\Gamma/dx_i$, our main purpose is to evaluate the distribution in the scaled-energy ($x_B$) of B-mesons  in the polarized top quark rest frame. For this study, we evaluate the partial decay width of process~(\ref{pros}) differential in $x_B$, $d\Gamma/dx_B$, at NLO where the normalized energy fraction of the B-meson is defined as $x_B=2E_B/(m_t(1-y))$.
According to the factorization  theorem of the QCD-improved parton model \cite{jc}, the energy distribution of a B-meson  can be expressed as the convolution of the parton-level spectrum with the nonperturbative fragmentation function $D_i^B(z, \mu_F)$,  describing the hadronization $i\rightarrow B$,
\begin{equation}\label{eq:master}
\frac{d\Gamma}{dx_B}=\sum_{i=b, g}\int_{x_i^\text{min}}^{x_i^\text{max}}
\frac{dx_i}{x_i}\,\frac{d\hat\Gamma}{dx_i}(\mu_R, \mu_F) D_i^B(\frac{x_B}{x_i}, \mu_F),
\end{equation}
where $d\hat\Gamma/dx_i (i=b, g)$ is the parton-level differential width  of the process (\ref{pros}) in each selected helicity coordinate system. In the equation above, $\mu_F$ and $\mu_R$ are the factorization and the renormalization scales, respectively. In principle, one can use two different values for these   scales; however, a choice often made consists of setting $\mu_R=\mu_F$ and we shall adopt this convention in our work. We will go back to the factorization theorem in Sec.~\ref{sec:three}, when our numerical analysis is presented.

In the next section, we present our analytic results for $d\hat\Gamma/dx_i  (i=b, g)$ at NLO in the helicity system~2.

\boldmath
\section{Analytic results for $d\hat\Gamma/dx_i$}
\label{sec:two}
\unboldmath

In this section we study the NLO radiative corrections to the partial decay width $t(\uparrow)\to b+H^+$  in the general 2HDM, where $H_1$ and $H_2$ are the doublets that their vacuum expectation values give masses  to the down and up type quarks, respectively, and a linear combination of the charged components of $H_1$ and $H_2$ gives the physical charged Higgs $H^\pm$. 
In general models with two Higgs doublets and generic coupling to all the quarks, it is difficult to avoid tree-level flavor-changing neutral currents. We, thus, limit ourselves to the models that naturally stop these problems by restricting the Higgs coupling to all quarks. Generally, there are two possibilities (called two models in the following) for the two Higgs doublets to couple to the fermions. 
In these models, the relevant part of the interaction Lagrangian is expressed as \cite{higg}
\begin{eqnarray}
L_I&=&\frac{g_W}{2\sqrt{2}m_W}V_{tb}H^+\big[\bar{u}_t(p_t)\{A(1+\gamma_5)+\nonumber\\
&&\hspace{2.5cm}B(1-\gamma_5)\}u_b(p_b)\big],
\end{eqnarray}
where $A$ and $B$ are model-dependent parameters and $g_W$ is the weak coupling factor.
In the first model (model $I$) the doublet $H_1$ couples  to all bosons and the doublet $H_2$ couples to all the quarks. In this model, one has
\begin{eqnarray}\label{model1}
A=m_t\cot\beta\quad , \quad B=-m_b\cot\beta.
\end{eqnarray}
In the second model (model $II$), the doublet $H_1$ couples to the right-chiral down-type quarks and the doublet $H_2$ couples to the right-chiral up-type quarks. In this model, the interaction Lagrangian consists of 
\begin{eqnarray}\label{model2}
A=m_t\cot\beta\quad , \quad B=m_b\tan\beta.
\end{eqnarray}
In (\ref{model1}) and (\ref{model2}), $\tan\beta=\textbf{v}_2/\textbf{v}_1$ is  the ratio of the vacuum expectation values of the two electrically neutral components of the two Higgs doublets. These models are also known as type-I and type-II 2HDM scenarios.

In the following, we express the technical detail of our calculation for the Born-term rate, the virtual and real gluon corrections.

\boldmath
\subsection{Born-level rate of $t\rightarrow bH^+$ in ZM-VFNS}
\unboldmath

The Born term amplitude in the MSSM for the process $t(\uparrow)\rightarrow b+H^+$ can either be expressed 
as a superposition of right- and left-chiral coupling factors, i.e. \textit{$M_0=\bar{u}_b\{g_t(1+\gamma_5)/2+g_b(1-\gamma_5)/2\}u_t$}, or 
as a superposition of  scalar and pseudoscaler coupling factors, i.e. \textit{$M_0=\bar{u}_b(a+b\gamma_5)u_t$},  where $a=(g_t+g_b)/2 $ and $b=(g_t-g_b)/2$. One also  has $g_b=2B$ and $g_t=2A$ where $A$ and $B$ are defined in (\ref{model1}) and (\ref{model2}) for the models I and II. The inverse relation reads $a=A+B$ and $b=A-B$.\\
Therefore, for the Born amplitude squared one has: $
|M_0|^2=2(p_b\cdot p_t)(a^2+b^2)+2(a^2-b^2)m_bm_t+4ab m_t(p_b\cdot s_t)$
where we replaced $\sum_{s_t}u(p_t, s_t)\bar{u}(p_t, s_t)=(\displaystyle{\not}{p}_t+m_t)$ in the unpolarized Dirac string by $u(p_t, s_t)\bar{u}(p_t, s_t)=(1-\gamma_5 \displaystyle{\not}{s}_t )(\displaystyle{\not}{p}_t+m_t)/2$ in the polarized state.\\
Considering Fig.~\ref{lo}, the polarization four-vector of the top quark in the top rest frame reads; $s_t=P(0;\sin\theta_P \cos\phi_P, \sin\theta_P\sin\phi_P,\cos\theta_P)$ and thus one has $p_b\cdot s_t=-P(|\vec{p}_b|\cos\theta_P)$. This justifies the minus sign in Eq.~(\ref{widthdefine}).
Therefore, the  tree-level helicity structure of differential rate reads
\begin{eqnarray}
\frac{d^2\hat\Gamma_0}{dx_b d\cos\theta_P}=\frac{1}{2}\bigg\{\hat\Gamma_0^{unpol}-P\hat\Gamma_0^{pol}\cos\theta_P\bigg\}\delta(1-x_b),
\end{eqnarray}
where the  unpolarized Born-level decay width is given by
\begin{eqnarray}
\hat\Gamma_0^{unpol}&=&\frac{m_t(a^2+b^2)}{16\pi }(1+R-y)\times\nonumber\\
&&\lambda^{\frac{1}{2}}(1,R,y)\bigg\{1+\frac{2\sqrt{R}}{1+R-y}(\frac{a^2-b^2}{a^2+b^2})\bigg\},
\end{eqnarray}
and  the polarized tree-level one, reads
\begin{eqnarray}\label{gammatree}
\hat\Gamma_0^{pol}&=&\frac{m_t}{8\pi}\lambda(1,R, y)(ab),
\end{eqnarray}
where  $\lambda(x,y,x)=x^2+y^2+z^2-2(xy+xz+yz)$ is the triangle function, $R=m_b^2/m_t^2$ and $y=m_{H^+}^2/m_t^2$.
The above results are independent of the selected helicity frames and are in complete agreement with Refs.~\cite{kadeer,Ali:2009sm,Czarnecki,Liud,Li:1990cp}.

In the limit of vanishing b-quark mass ($m\to 0\equiv R\to 0$) one has  $a=b$ in the model $I$ (or in the type-I 2HDM), then the tree-level  decay width is simplified to
\begin{eqnarray}\label{gamma}
\hat\Gamma_0^{unpol}=\hat\Gamma_0^{pol}=\frac{m_t^3}{8\sqrt{2}\pi}G_F|V_{tb}|^2(1-y)^2\cot^2\beta,
\end{eqnarray}
and for the model $II$ (type-II 2HDM), one has
\begin{eqnarray}\label{khoram}
\hat\Gamma_0^{unpol}&=&\frac{m_t^3}{8\sqrt{2}\pi}G_F|V_{tb}|^2(1-y)^2\big\{\cot^2\beta+R\tan^2\beta\big\}\times\nonumber\\
&&\bigg(1+\frac{4R}{1-y}(\frac{1}{\cot^2\beta+R\tan^2\beta})\bigg),
\end{eqnarray}
and,
\begin{eqnarray}\label{khorami}
\hat\Gamma_0^{pol}&=&\frac{m_t^3}{8\sqrt{2}\pi}G_F|V_{tb}|^2(1-y)^2
\big\{\cot^2\beta-R\tan^2\beta\big\}.\nonumber\\
 \end{eqnarray}
In (\ref{khoram}) and (\ref{khorami}), when the $R\tan^2\beta$-term  can be compared with   $\cot^2\beta$  therefore one cannot naively set $m_b=0$ in all expressions. 
For example, if we take $m_b=4.78$~GeV, $m_t=172.98$~GeV and from the unexcluded regions of the MSSM $m_{H^+}-\tan\beta$ parameter space \cite{CMS:2014cdp, TheATLAScollaboration:2013wia}
we also take $m_{H^+}=155$~GeV and $\tan\beta=4$ thus the second term in the curly brackets in (\ref{khoram}) and (\ref{khorami}) can become as large as ${\cal O}(20\%)$ and this order will be larger when $\tan\beta$ is increased. Therefore, the $m_b\to 0$ approximation is not suitable for the type-II 2HDM. In this paper we work in the type-I 2HDM and adopt, with a very good approximation,  the Born term presented in (\ref{gamma}) in the massless or zero-mass variable-flavor-number (ZM-VFN) scheme \cite{Binnewies:1998vm} where the zero mass parton approximation is also applied to the bottom quark and the nonzero value of the b-quark mass only enter through the initial condition of the nonperturbative FF. 
 
In the following,  we present our analytical results for the ${\cal O}(\alpha_s)$ QCD corrections to the tree-level decay rate  in the ZM-VFN scheme.

\boldmath
\subsection{Virtual Corrections}\label{virtual}
\unboldmath

The QCD virtual one-loop  corrections to the polarized differential width arise from  emission and absorption of a virtual gluon from the same quark leg (quark self-energy) and from a virtual gluon exchanged between the top and bottom quark legs (vertex correction). In the ZM-VFN scheme all divergences  including the  infra-red (IR) and ultra-violet (UV) singularities which arise from the collinear- and the soft-gluon  emissions, respectively,  are regularized by dimensional regularization in $D=4-2\epsilon$ space-time dimensions to become single poles in $\epsilon$. These singularities are subtracted at factorization scale $\mu_F$ and absorbed into the bare FFs according to the modified minimal-subtraction  scheme ($\overline{MS}$). The virtual contributions are the same in both helicity systems 1 and 2, and more detail of our calculation can be found in \cite{MoosaviNejad:2016aad}. We just mention that by neglecting  the b-quark mass the counter term of the vertex consists of the top quark mass renormalization and the wave function renormalizations of both top and bottom quarks. Here, we just  present our final result of the virtual corrections to the polarized differential decay rate as
\begin{eqnarray}
\frac{d\hat\Gamma^{vir,pol}}{dx_b}&=&\hat\Gamma_0^{pol}\frac{\alpha_s(\mu_R)}{2\pi}C_F\delta(1-x_b)\big(-\frac{1}{\epsilon^2}+\frac{F}{\epsilon}-\frac{F^2}{2}\nonumber\\
&&+(\frac{2}{y}-5)\ln(1-y)-2Li_2(y)-\frac{7}{8}-\frac{\pi^2}{12}\big),\nonumber\\
\end{eqnarray}
where, $F=2\ln(1-y)-\ln(4\pi \mu_F^2/m_t^2)+\gamma_E-5/2$, $C_F=(N_c^2-1)/(2N_c)=4/3$ for $N_c=3$ quark colors,  and $Li_2(x)=-\int_0^x(dt/t)\ln(1-t)$ is the Spence function. Note that, all UV-divergences  are canceled after summing all virtual corrections up but the IR-singularities are remaining which are labeled  by $\epsilon$ in the above equation. Since the virtual corrections are the same in both helicity systems 1 and 2, the above result is in agreement with \cite{kadeer} where the authors have considered the first helicity system.

\boldmath
\subsection{Real gluon Corrections}\label{real}
\unboldmath

In this section we calculate the  ${\cal O}(\alpha_s)$ QCD corrections (i.e. $t(\uparrow)\to bH^+g$) which are needed  to cancel the IR-singularities of the virtual corrections. 
In the rest frame of a top quark decaying into a Higgs boson, a bottom quark and a gluon the outgoing particles define an event plane so that relative to this plane one can define the spin direction of the polarized top quark. For our aim, two possible coordinate systems are defined as
\begin{eqnarray}
Systsem~1&:& \vec{p}_{H^+} ||\hat{z}; (\vec{p}_b)_x\geq 0\nonumber\\
Systsem~2&:& \vec{p}_b ||\hat{z}; (\vec{p}_{H^+})_x\geq 0
\end{eqnarray}
The various helicity systems provide independent probes of light charged Higgs  bosons in the polarized top quark decay dynamics.

In \cite{MoosaviNejad:2016aad}, we analyzed   the spin-momentum correlation between the top quark polarization vector and the momenta of its decay products in the system~1. In the present work, we study the same analysis in the system~2 and show that the energy spectrum of the outgoing B-meson depends on the helicity system selected. Considering the general form of angular distribution of the differential decay width (\ref{widthdefine}), one has 
\begin{eqnarray}\label{real}
\frac{d^2\hat\Gamma^{real}}{dx_b d\cos\theta_P}=\frac{1}{2}(\frac{d\hat\Gamma^{unpol,real}}{dx_b}-P\frac{d\hat\Gamma^{pol,real}}{dx_b}\cos\theta_P),
\end{eqnarray}
where $d\hat\Gamma^{unpol}/dx_b$ is presented in \cite{MoosaviNejad:2011yp}.

The ${\cal O}(\alpha_s)$  real gluon (tree-graph) contribution to the differential decay rate results from the square of the real amplitude as  $|M^{\textbf{real}}|^2=M^{\textbf{real} \dagger}\cdot M^{\textbf{real}}$, where $M^{\textbf{real}}$  reads
\begin{eqnarray}\label{finfin}
M^{\textbf{real}}&=&g_s\frac{\lambda^a}{2}\bar u(p_b, s_b)\big\{\frac{2p_t^\mu-
	\displaystyle{\not}p_g \gamma^\mu}{2p_t \cdot p_g}
\\
&&-\frac{2p_b^\mu+\gamma^\mu \displaystyle{\not}p_g}
{2p_b\cdot p_g}\big\}(a\textbf{1}+b\gamma_5) u(p_t, s_t)\epsilon_{\mu}^{\star}(p_g,r),\nonumber
\end{eqnarray}
where the polarization vector of the real gluon with the momentum $p_g$ and spin $r$ is denoted by $\epsilon(p_g,r)$. The first and second terms in the curly brackets refer to the real gluon emission from the top  and the bottom quarks, respectively.

As before, to regulate the IR-divergences we work in $D=4-2\epsilon$  dimensions, therefore from the definition of decay rate, one has
\begin{eqnarray}\label{decaydef}
d\hat\Gamma^{real}=\frac{\mu_F^{2(4-D)}}{2m_t}{|M^{real}|^2}dPS(p_t, p_b, p_g, p_{H^+}),
\end{eqnarray}
where, the Phase Space element reads
\begin{eqnarray}
dPS&=&\frac{d^{D-1}\bold{p}_b}{(2\pi)^{D-1}2E_b}\frac{d^{D-1}\bold{p}_{H^+}}{(2\pi)^{D-1}2E_{H^+}}\frac{d^{D-1}\bold{p}_g}{(2\pi)^{D-1}2E_g}\nonumber\\
&&\times(2\pi)^D\delta^D(p_t-p_b-p_{H^+}-p_g).
\end{eqnarray}
To calculate the real polarized differential decay rate $d\hat\Gamma^{pol,real}/dx_b$,  we fix the momentum of the bottom quark in Eq.~(\ref{decaydef}) and integrate over the gluon energy  which ranges as $m_tS(1-x_b)\leq E_g\leq m_t S(1-x_b)/(1-2 S x_b)$ where $S=(1-y)/2$. Also, to get the correct finite terms one has to normalize it to the Born width (\ref{gamma}) which is evaluated in the dimensional regularization at ${\cal O}(\epsilon^2)$, i.e.
$\hat\Gamma_0^{pol}\rightarrow \hat\Gamma_0^{pol}\{1-\epsilon (\gamma_E+2\ln S-\ln(4\pi\mu_F^2/m_t^2))\}$.\\
Thus, in the second helicity coordinate system the contribution of the real gluon emission into the normalized differential decay width is  given by
\begin{eqnarray}
\frac{1}{\hat\Gamma_0^{pol}}\frac{d\hat\Gamma^{real,pol}}{dx_b}&=&\frac{\alpha_s}{2\pi}C_F\Big\{\delta(1-x_b)\big[\frac{1}{\epsilon^2}-\frac{1}{\epsilon}(F+\frac{3}{2})+\frac{F^2}{2}\nonumber\\
&&+\frac{3}{2}F-2\frac{y}{1-y}\ln y+2Li_2(1-y)-\frac{\pi^2}{4}\nonumber\\
&&+\frac{5}{8}\big]+\frac{1+x_b^2}{(1-x_b)_+}\big[-\frac{1}{\epsilon}+2\ln x_b+F\nonumber\\
&&+\frac{3}{2}\big]+2(1+x_b^2)\bigg(\frac{\ln(1-x_b)}{1-x_b}\bigg)_+\Big\},
\end{eqnarray}
where $F=2\ln(1-y)-\ln(4\pi \mu_F^2/m_t^2)+\gamma_E-5/2$ and the plus distributions are defined as usual.

\boldmath
\subsection{Analytic Results for Partial Decay Rates $d\hat\Gamma/dx_i$ in the helicity system~2}
\unboldmath

The NLO expression for the $d\hat\Gamma^{pol}/dx_b$ is obtained by summing the Born term, the virtual one-loop and the real gluon contributions. Our  result for the  helicity coordinate system~2, is as follows
\begin{eqnarray}\label{fin}
\frac{d\hat\Gamma^{pol}}{dx_b}&=&\hat\Gamma_0^{pol}\Bigg\{\delta(1-x_b)+\frac{\alpha_s(\mu_R)}{2 \pi}\bigg\{\big[-\frac{1}{\epsilon}+\gamma_E-\ln 4\pi\big]\nonumber\\
&&\times P_{qq}^{(0)}(x_b)+C_F\Big[\delta(1-x_b)\big[2\frac{1-y}{y}\ln(1-y)\nonumber\\
&&-4-\frac{2y}{1-y}\ln y-\frac{\pi^2}{3}-2Li_2(y)+2Li_2(1-y)\nonumber\\
&&-\frac{3}{2}\ln\frac{\mu_F^2}{m_t^2}\big]-\frac{1+x_b^2}{(1-x_b)_+}\big[1-2\ln(x_b(1-y))\nonumber\\
&&+\ln\frac{\mu_F^2}{m_t^2}\big]+2(1+x_b^2)\bigg(\frac{\ln(1-x_b)}{1-x_b}\bigg)_+\Big]\bigg\}\Bigg\},
\end{eqnarray}
where  $P_{qq}^{(0)}$ is the time-like $q\to q$ splitting function at leading order \cite{dglap}, so
\begin{eqnarray}\label{eq:zmvfn}
P_{qq}^{(0)}(x_b) = C_F \bigg(\frac{1+x_b^2}{(1-x_b)_+}+\frac{3}{2}\delta(1-x_b)\bigg).
\end{eqnarray}

Since, the bottom-flavored hadrons can be also produced through the hadronization of the emitted real gluon at NLO,  we also need the differential decay rate $d\hat\Gamma^{pol}/dx_g$ in the ZM-VFN scheme. To calculate the $d\hat\Gamma^{pol}/dx_g$ we start form Eq.~(\ref{decaydef}) and fix the momentum of the gluon and integrate over the bottom quark energy so that $m_t S(1-x_g)\leq E_b\leq m_t S(1-x_g)/(1-2 S x_g)$. Since we fix the gluon momentum, then there will be no soft singularities in the $d\hat\Gamma^{pol}/dx_g$. The result in the helicity system~2, reads
\begin{eqnarray}\label{finny}
\frac{d\hat\Gamma^{pol}}{dx_g}&=&\hat\Gamma_0^{pol}\Bigg\{\frac{\alpha_s(\mu_R)}{2\pi}\bigg\{\big[-\frac{1}{\epsilon}+\gamma_E-\ln4\pi\big]\times P_{gq}^{(0)}(x_g)\nonumber\\
&&+C_F\Big[3-\frac{y^2}{4S(1-2Sx_g)^2}+\frac{1}{Sx_g^2}\ln(1-2Sx_g)\nonumber\\
&&+\frac{12S^2-8S+1}{4S(1-2Sx_g)}-\frac{x_g}{2}-\frac{1+(1-x_g)^2}{x_g}\bigg(\ln\frac{\mu_F^2}{m_t^2}\nonumber\\
&&-\ln\frac{4S^2x_g^2(1-x_g)^2}{1-2Sx_g}\bigg)\Big]\bigg\}\Bigg\},
\end{eqnarray}
where  $P_{gq}^{(0)}$ is the time-like $q\to g$ splitting function at LO \cite{dglap},
\begin{eqnarray}\label{eq:zmvfn}
P_{gq}^{(0)}(x_g) &=& C_F \big(\frac{1+(1-x_g)^2}{x_g}\big).
\end{eqnarray}

To subtract the collinear singularities remaining in Eqs.~(\ref{fin}) and (\ref{finny}), we apply  the modified minimal subtraction ($\overline{MS}$) scheme where the collinear singularities are absorbed into the bare FFs. This renormalizes the FFs and generates the finite terms of the form $\alpha_s\ln(m_t^2/\mu_F^2)$ in the polarized differential decay rates. According to this scheme, in order to  get the $\overline{MS}$  coefficient functions we shall have to subtract from Eqs.~(\ref{fin}) and (\ref{finny}) the ${\cal O}(\alpha_s)$ term multiplying the  characteristic $\overline{MS}$ constant $(-1/\epsilon+\gamma_E-\ln 4\pi)$. 
In this work we set $\mu_R=\mu_F=m_t$, so that in Eqs.~(\ref{fin}) and (\ref{finny}) the terms proportional to $\ln(m_t^2/\mu_F^2)$ vanish.

Integrating $d\hat\Gamma^{pol}/dx_b$ of Eq.~(\ref{fin}) over $x_b(0<x_b<1)$, we obtain the NLO renormalized decay rate as
\begin{eqnarray}\label{rate}
\hat\Gamma^{pol}&=&\hat\Gamma_0^{pol}\Big\{1-\frac{C_F\alpha_s}{2\pi}\big[\frac{2y}{1-y}\ln y+(5-\frac{2}{y})\ln(1-y)+\nonumber\\
&&2Li_2(y)-2Li_2(1-y)-\frac{7}{2}+\pi^2\big]\Big\}.
\end{eqnarray}
Our previous result for $\hat\Gamma^{pol}$($=\int_0^1 dx_b d\hat\Gamma^{pol}/dx_b$) in the  helicity system~1 \cite{MoosaviNejad:2016aad} was in complete agreement with Ref.~\cite{kadeer}, but the above result computed in the second frame (system~2) is completely new.

\section{Numerical analysis in  type-I 2HDM}
\label{sec:three}

In the MSSM, the mass of charged Higgs bosons  is restricted  by $m_{H^\pm}>m_{W^{\pm}}$ at tree-level \cite{Nakamura:2010zzi}, but this restriction does not hold for some regions of parameter space after including radiative corrections.
In this model, $m_{H^\pm}$ is strongly correlated with the mass of other Higgs bosons.
In \cite{Ali:2009sm}, it is mentioned that 
a charged Higgs  boson with a mass  range $80 GeV\leq m_{H^\pm}\leq 160 GeV$ is a logical possibility
and its effects should be searched for in the decay mode $t\rightarrow  bH^+\rightarrow B\tau^+\nu_\tau+X$.
On the other hand, the recent results of a search for evidence of a charged Higgs boson in $19.5-19.7 fb^{-1}$ of proton-proton collision data recorded at $\sqrt{s}=8$~TeV are reported by the CMS \cite{CMS:2014cdp} and the ATLAS \cite{TheATLAScollaboration:2013wia}  collaborations, using the $\tau+jets$ channel with a hadronically decaying $\tau$ lepton in the final state. 
According to Fig.~7 of Ref.~\cite{TheATLAScollaboration:2013wia}, the large region in the MSSM $m_{H^+}-\tan\beta$ parameter space is excluded for $m_{H^+}=80-160$~GeV. So, the unexcluded regions of this parameter space include the charged Higgs masses as $90\leq m_{H^+}\leq 100$~GeV (with $6<\tan\beta <10$) and  $140\leq m_{H^+}\leq 160$~GeV (with $3<\tan\beta<21$). See also Fig.~9 of Ref.~\cite{CMS:2014cdp}.
Therefore, these values of  $m_{H^\pm}$ and $\tan\beta$ are still allowed and in this work our prediction and analysis is restricted to these regions. However, a definitive search of the charged Higgs bosons over this part of the $m_{H^+}-\tan\beta$  parameter space  is a program that still has to be carried out and this belongs to the LHC experiments. 

Here, for our numerical analysis we adopt the input parameter values from Ref.~\cite{Nakamura:2010zzi} as; 
$G_F = 1.16637\times10^{-5}$~GeV$^{-2}$,
$m_t = 172.98$~GeV,
$m_b=4.78$~GeV,
$m_W=80.399$~GeV,
$m_B = 5.279$~GeV, and
$|V_{tb}|=0.999152$.
Considering the unexcluded $m_{H^+}-\tan\beta$ parameter space from the ATLAS experiments \cite{TheATLAScollaboration:2013wia}, we also consider $m_{H^+}=95, 155$~GeV and $160$~GeV. \\
In the ZM-VFN scheme the polarized and unpolarized decay rates at the Born level are the same in both helicity systems (see (\ref{gamma}) and  corresponding explanation) and from now we label them by $\hat\Gamma_0(=\hat\Gamma_0^{unpol}=\hat\Gamma_{1,0}^{pol}=\hat\Gamma_{2,0}^{pol})$. Our result for the unpolarized rate at NLO, is
\begin{eqnarray}
\hat\Gamma^{unpol}&=& \hat\Gamma_0 (1-0.010)\quad,\quad \textrm{for}\quad m_{H^+}=160 \textrm{GeV}\nonumber\\
\hat\Gamma^{unpol}&=& \hat\Gamma_0^{\prime} (1-0.028)\quad,\quad \textrm{for}\quad m_{H^+}=155 \textrm{GeV}\nonumber\\
\hat\Gamma^{unpol}&=& \hat\Gamma_0^{\prime\prime} (1-0.086)\quad,\quad \textrm{for}\quad m_{H^+}=95 \textrm{GeV}\nonumber
\end{eqnarray}
and for the polarized rate in the  system~1, one has
\begin{eqnarray}
\hat\Gamma^{pol}_1&=& \hat\Gamma_0 (1-0.037)\quad,\quad \textrm{for}\quad m_{H^+}=160 \textrm{GeV}\nonumber\\
\hat\Gamma^{pol}_1&=& \hat\Gamma_0^{\prime} (1-0.038)\quad,\quad \textrm{for}\quad m_{H^+}=155 \textrm{GeV}\nonumber\\
\hat\Gamma^{pol}_1&=& \hat\Gamma_0^{\prime\prime} (1-0.043)\quad,\quad \textrm{for}\quad m_{H^+}=95 \textrm{GeV}\nonumber
\end{eqnarray}
and for the NLO polarized width  in the helicity system~2, we have
\begin{eqnarray}
\hat\Gamma^{pol}_2&=& \hat\Gamma_0 (1-0.033)\quad,\quad \textrm{for}\quad m_{H^+}=160 \textrm{GeV}\nonumber\\
\hat\Gamma^{pol}_2&=& \hat\Gamma_0^{\prime} (1-0.051)\quad,\quad \textrm{for}\quad m_{H^+}=155 \textrm{GeV}\nonumber\\
\hat\Gamma^{pol}_2&=& \hat\Gamma_0^{\prime\prime} (1-0.109)\quad,\quad \textrm{for}\quad m_{H^+}=95 \textrm{GeV}\nonumber
\end{eqnarray}
In the above results the $\hat\Gamma_0$, $\hat\Gamma_0^{\prime}$ and $\hat\Gamma_0^{\prime\prime}$ depend on the $m_{H^+}$ and $\tan\beta$, see (\ref{gamma}).
%By defining the polarization asymmetry as  $\alpha_H=\Gamma^{pol}/\Gamma^{unpol}$, for model 1 one has
%\begin{eqnarray}
%\alpha_H=\frac{1-R}{(1-R)^2-y(1+R)}\lambda^{\frac{1}{2}}(1,R,y),
%\end{eqnarray}
%which does not depend on $\tan\beta$, and for model 2 the result is
%\begin{eqnarray}
%\alpha_H=\frac{\cot^2\beta-R\tan^2\beta}{(1+R-y)(\cot^2\beta+R\tan^2\beta)+4R}\lambda^{\frac{1}{2}}(1,R,y).\nonumber\\
%\end{eqnarray}
As is seen, the NLO polarized decay rates depend on the selected helicity coordinate system. 
In obtaining the results above, we applied the unpolarized decay rate presented in \cite{MoosaviNejad:2011yp} and the polarized ones in the first and second helicity systems given in \cite{MoosaviNejad:2016aad} and (\ref{rate}), respectively. 

After our numerical analysis of decay widths we are now in a situation to present our phenomenological predictions 
for the scaled-energy ($x_B$) spectrum  of bottom-flavored mesons (B) inclusively  produced in  polarized top decays in the type-I 2HDM.
To show  our predictions for the $x_B$-distribution, we
consider the doubly differential distribution  $d^2\Gamma/(dx_B d\cos\theta_P)$ of the partial width of the decay $t(\uparrow)\rightarrow BH^++X$ in the system~2. Here,  
$x_B=2E_B/(m_t(1-y))$ is the scaled-energy fraction of the B-meson in the top quark rest frame, where the energy of B-meson  ranges from $E_B^{min}=m_B$ to $E_B^{max}=(m_t^2+m_B^2-m_H^2)/(2m_t)$.\\
According to the factorization formula (\ref{eq:master}), the required ingredients for this study are the parton-level differential decay widths (\ref{fin}) and (\ref{finny}) and the fragmentation functions (FFs) $D_b^B(z)$ and $D_g^B(z)$ which describe the splitting of $b\to B$ and $g\to B$, respectively.
To describe these hadronization processes, from Ref.~\cite{Kniehl:2008zza} we employ the nonperturbative $B$-hadron  FFs determined at NLO in the ZM-VFN scheme through a global fit  to
$e^+e^-$ annihilation data taken by OPAL
\cite{Abbiendi:2002vt}, ALEPH \cite{Heister:2001jg} and SLD \cite{Abe:1999ki}. In Ref.~\cite{Kniehl:2008zza} authors used  a simple power model  $D_b(z,\mu_F^\text{ini})=Nz^\alpha(1-z)^\beta$
as the initial condition for the $b\to B$ FF at $\mu_F^\text{ini}=4.5$~GeV, while the  gluon and light-quark FFs were generated
via the DGLAP  evolution equations \cite{dglap}.
The fit yielded the values $N=4684.1$, $\alpha=16.87$, and $\beta=2.628$  for the FF parameters.
%\begin{figure}
%	\begin{center}
%		\includegraphics[width=0.7\linewidth,bb=110 32 490 540]{plot}
%		\caption{\label{plot}%
			%Exclusion region in the MSSM $\tan\beta-m_{H^+}$ parameter %space for $m_{H^+}=80-160$ GeV is shown. The $\pm 1\sigma$ %and $\pm 2\sigma$ bands around the expected limit are also %shown. The light grey region is excluded. Plot is got from %Ref.~\cite{CMS:2014cdp}.}
%	\end{center}
%\end{figure}
\begin{figure}
	\begin{center}
		\includegraphics[width=0.7\linewidth,bb=137 42 690 690]{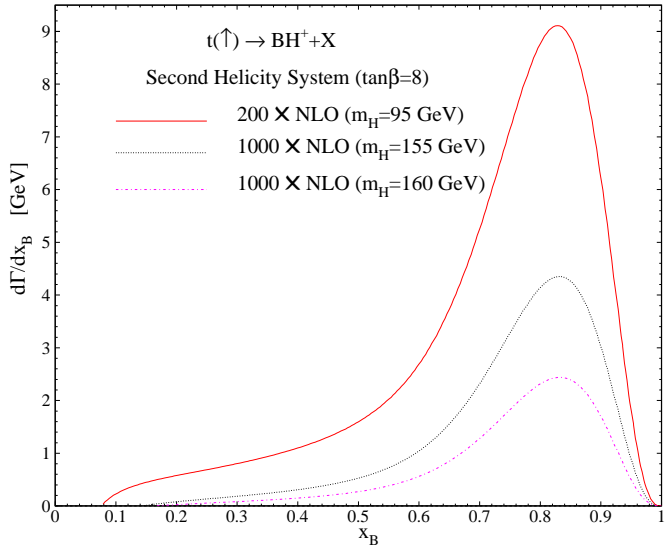}
		\caption{\label{fig1}%
			The NLO $x_B$-spectrum ($d\Gamma/dx_B$) in polarized top decay in the helicity coordinate system~2 with $\tan\beta=8$  and $m_{H^+}=95$~GeV (solid line), $m_{H^+}=155$ (dotted line) and $160$~GeV (dot-dashed line). }
	\end{center}
\end{figure}
\begin{figure}
	\begin{center}
		\includegraphics[width=0.7\linewidth,bb=137 42 690 690]{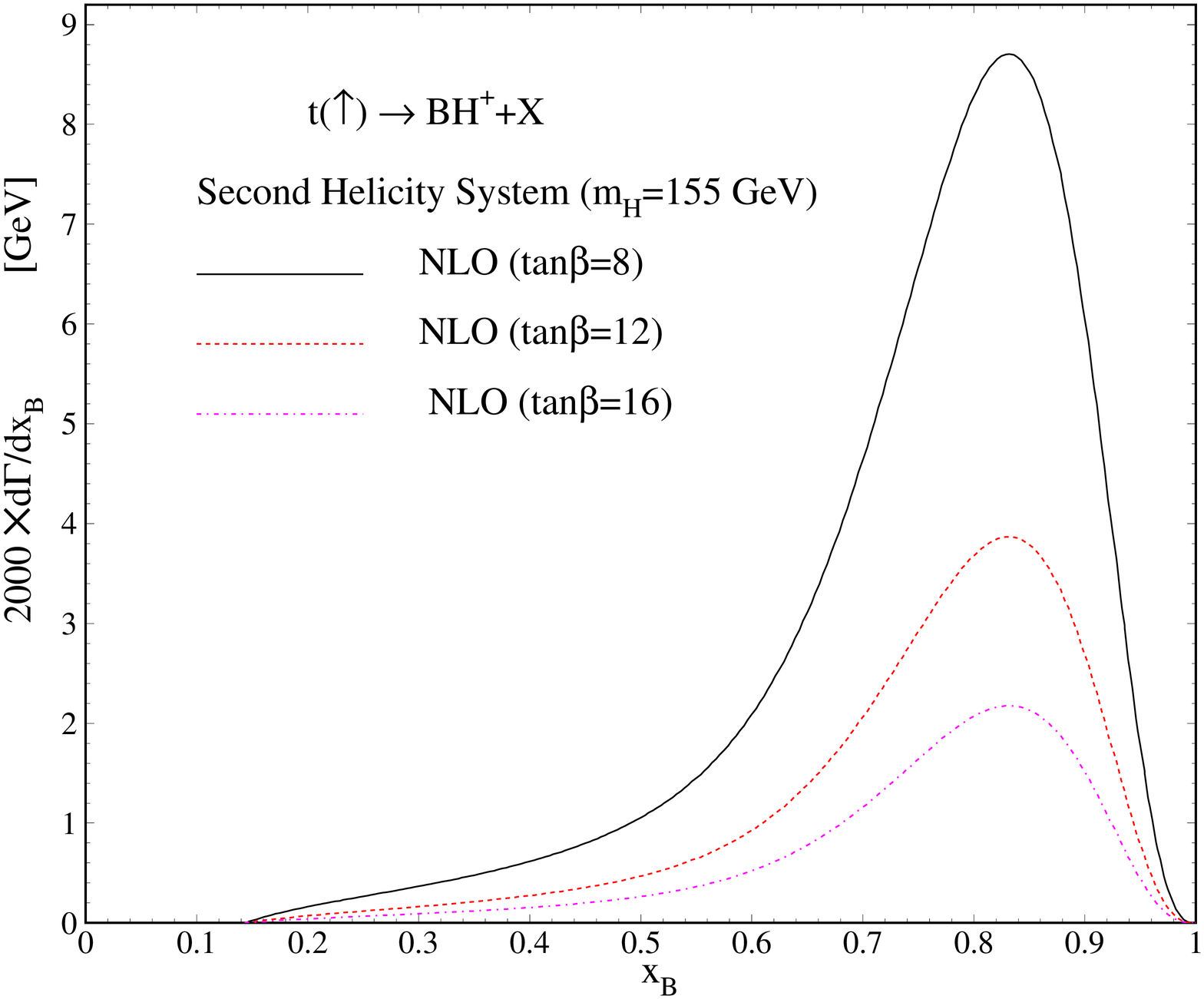}
		\caption{\label{fig2}%
			$x_B$ spectrum  in polarized top decay in the type-I 2HDM  with different values of $\tan\beta=8$, $12$ and $16$. The charged Higgs boson mass is set to $m_{H^+}=155$~GeV. Analysis is done in the second helicity coordinate system.}
	\end{center}
\end{figure}
\begin{figure}
	\begin{center}
		\includegraphics[width=0.7\linewidth,bb=137 42 690 690]{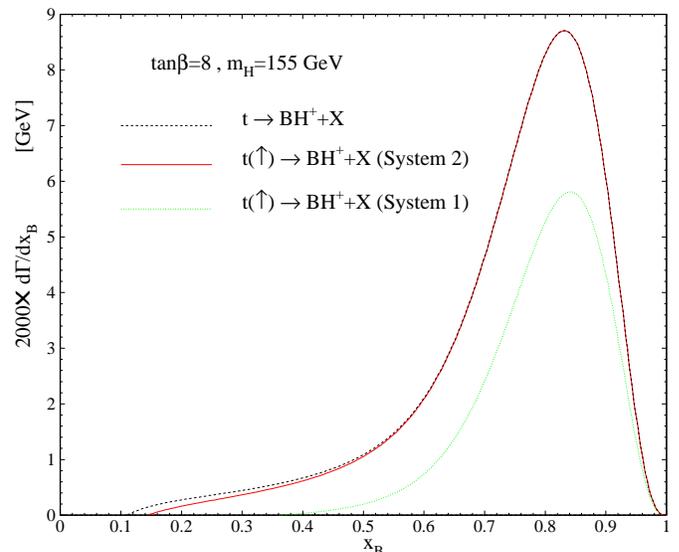}
		\caption{\label{fig3}%
			$d\Gamma/x_B$ as a function of $x_B$ in the type-I 2HDM considering the ZM-VFN  scheme. The  unpolarized (dashed line) and polarized partial decay rates are compared at NLO  taking $m_{H^+}=155$~GeV, $\tan\beta=8$ and $\mu_R=\mu_F=m_t$. For the polarized top decays we used the helicity system~1 (dotted line) and 2 (solid line). Details are discussed in the text.}
	\end{center}
\end{figure}
\begin{figure}
	\begin{center}
		\includegraphics[width=0.7\linewidth,bb=137 42 690 690]{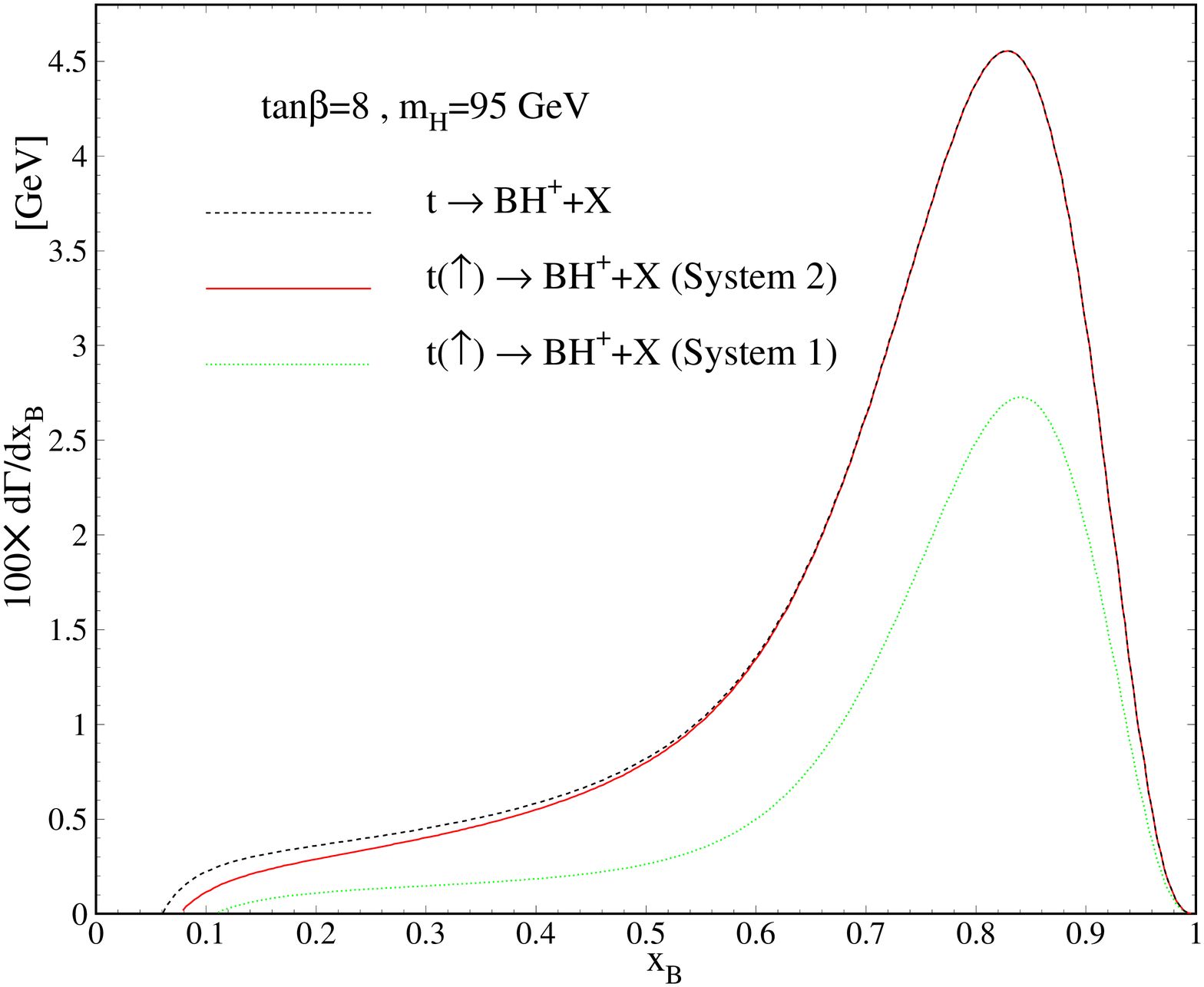}
		\caption{\label{fig4}%
			As in Fig.~\ref{fig3}, but for $m_{H^+}=95$~GeV. This mass is not excluded by the ATLAS  experiments \cite{TheATLAScollaboration:2013wia}.}
	\end{center}
\end{figure}

Considering the unexcluded $m_{H^+}-\tan\beta$ parameter space from 
the CMS \cite{CMS:2014cdp} and the ATLAS \cite{TheATLAScollaboration:2013wia} experiments, in Fig.~\ref{fig1} we show our prediction for the $x_B$-spectrum at NLO in the system~2, taking  $m_{H^+}=95$~GeV (solid line),  $m_{H^+}=155$~GeV (dotted line) and $m_{H^+}=160$~GeV (dot-dashed line) where $\tan\beta=8$ is fixed for all predictions. As is seen, when $m_{H^+}$  increases the size of decay rate decreases but the peak position is shifted towards higher values of $x_B$.

Considering the results of the CMS \cite{CMS:2014cdp} and  ATLAS \cite{TheATLAScollaboration:2013wia} experiments where $4\leq \tan\beta\leq 16$ is  allowed for $m_{H^+}=155$~GeV, in Fig.~\ref{fig2}  we study the energy spectrum of  B-meson in the helicity system~2 for 
different values of $\tan\beta=8$ (solid line), $12$ (dashed line) and $16$ (dot-dashed line), where the mass of Higgs boson is set to $m_{H^+}=155$~GeV for all analysis. As is seen, when $\tan\beta$  increases the size of decay rate decreases. This is obvious because  $\hat\Gamma_0$ (\ref{gamma}) is proportional to $\cot^2\beta$. 

In Fig.~\ref{fig3}, taking $m_{H^+}=155$~GeV and $\tan\beta=8$ the NLO energy spectrum of B-mesons from polarized top decays, $t(\uparrow)\rightarrow BH^++X$, in the first (dotted line) and second (solid line) helicity coordinate systems are shown. As is seen the energy distributions obtained from our analysis in two various systems are  different and the size of NLO correction is larger in the system~2. For more comparison, we also plotted the energy distribution of B-mesons through unpolarized top quark decays (dashed line). A considerable point is that the size of  NLO corrections is the same both for the polarized top decay in the helicity system~2 and for the unpolarized one, except for small values of $x_B (0.14<x_B<0.45)$. In this region the unpolarized distribution is larger. \\
In Fig.~\ref{fig4}, as in Fig.~\ref{fig3}, the same comparisons  are done but for $m_{H^+}=95$~GeV.
Our results show that in these cases the NLO corrections are similar in the shape, however, the unpolarized distribution shows a more enhancement in size at NLO.

It should be pointed out that our formalism elaborated here can be also extended  to the production of hadron species other than bottom-flavored hadrons, such as pions, kaons and protons, etc.,  using the nonperturbative $(b, g)\rightarrow \pi/K/P$ FFs extracted in our recent works \cite{Soleymaninia:2013cxa,Nejad:2015fdh},
relying on their universality and scaling violations \cite{collins}.

\section{Conclusions}
\label{sec:four}

Charged Higgs bosons ($H^\pm$) are predicted in models consisting of at least  two Higgs doublets, of which the simplest are the two-Higgs-doublet models (2HDM). 
The charged-Higgses have been searched for in high energy experiments, in particular, at the Tevatron, ATLAS and CMS but they have  not been seen so far.
The discovery of a charged Higgs would represent unambiguous evidence  for the presence of physics beyond the SM. There are many reasons, both from theoretical considerations and experimental observations, to except physics beyond the SM, such as the hierarchy problem, neutrino masses and dark matter.\\
In the 2HDM, the main production mode of light charged Higgs boson ($m_{H^+}<m_t$) is through the top quark decay, $t\to bH^+$. On the other hand, bottom quarks hadronize, via $b\to B+X$, before they decay, so that the decay process $t\to BH^++X$ is of prime importance at the LHC. Therefore, the study of scaled-energy ($x_B$) distribution of the bottom-flavored  mesons ($B$)  inclusively produced in  top quark decays is proposed as a new way to search for the light charged Higgs bosons. 
For this study, we need to evaluate the quantity $d\Gamma/dx_B$. 

In \cite{MoosaviNejad:2011yp}, we studied  the energy spectrum of the B-mesons in unpolarized top  decays into a charged-Higgs boson and a b-quark at NLO in the 2HDM. In \cite{MoosaviNejad:2016aad} we studied the spin-dependent energy distribution of B-mesons produced through the polarized top  decays  at NLO in a special helicity coordinate system (system~1), where the event plane lied in the $(x, z)$ plane and the Higgs three-momentum was along the $z$-axis. 
In the present work, we have presented results on the NLO radiative corrections to the spin-dependent differential width $d^2\Gamma/(dx_B d\cos\theta_P)$, applying a different helicity system (system~2) where the $z$-axis is defined by the b-quark 3-momentum. This provides an independent probe of charged Higgses. To make these predictions we obtained  the analytical results for the parton-level differential decay width $d\hat\Gamma(t(\uparrow)\to bH^+(+g))/dx_a (a=b, g)$ in two helicity systems~1 and 2. Our result for the unpolarized differential decay width $d\hat\Gamma(t\to bH^+)/dx_b$  was in complete agreement with Refs.~\cite{Ali:2009sm,Czarnecki,Liud,Li:1990cp} after integration over $0\leq x_b\leq 1$, and our result for the polarized one in the system~1 was in agreement with \cite{kadeer}  after integration over $x_b$.  Here, using the same techniques we calculated the polarized differential width in the helicity system~2 and we also computed, for the first time, the polarized rate in the system~2. 
We found that the polarized results depend on the selected helicity system, extremely.

For our numerical analysis, considering the recent results reported by the CMS \cite{CMS:2014cdp} and ATLAS \cite{TheATLAScollaboration:2013wia} collaborations we restricted ourselves to the unexcluded regions of the MSSM  $m_{H^+}-\tan\beta$ parameter space which include 
$90\leq m_{H^+}\leq 100$~GeV (with $6<\tan\beta <10$) and  $140\leq m_{H^+}\leq 160$~GeV (with $3<\tan\beta<21$). 

Since, highly polarized top quarks will become available at hadron colliders
through single top production processes, which occur at the $33\%$ level of the $t\bar{t}$ pair production rate \cite{Mahlon:1996pn},
and in top quark pairs produced in future linear $e^+e^-$-colliders \cite{Kuhn:1983ix} these studies can be considered as a new channel to search for the charged Higss bosons.
%The top quark polarization is studied by the angular correlations between the top quark spin and its decay products momenta, so these spin-momenta correlations will allow the detailed studies of the top decay mechanism in the 2HDM. In our previous work \cite{MoosaviNejad:2011yp},   we studied the energy spectrum of B-meson in the 2HDM for the unpolarized decay mode. Here, we also compared the energy spectrum of B-mesons produced both through the unpolarized and polarized top decays. Results show a considerable difference between two distributions, however, they depend on the charged Higgs mass and $\tan\beta$.

\section{Acknowledgments}
\label{sec7}
We would like to thank  the LHC top working group  for importance discussion and comments. We warmly acknowledge the CERN TH-PH division for its hospitality where a portion of this work was performed.


\begin{thebibliography}{25}
	
\bibitem{higg}
J.~F.~Gunion, H.~Haber, G.~Kane, and S.~Dawson,\textit{ The Higgs Hunter's Guide} (Addison-Wesley, Reading, MAA, 1990), and refrences therein.	
		
	
%\cite{Cheng:1980qt}
\bibitem{Cheng:1980qt}
T.~P.~Cheng and L.~F.~Li,
%``Neutrino Masses, Mixings and Oscillations in SU(2) x U(1) Models of Electroweak Interactions,''
Phys.\ Rev.\ D {\bf 22} (1980) 2860.
% doi:10.1103/PhysRevD.22.2860
%%CITATION = doi:10.1103/PhysRevD.22.2860;%%
%737 citations counted in INSPIRE as of 09 Nov 2016	
	
%\cite{Lee:1973iz}
\bibitem{Lee:1973iz}
T.~D.~Lee,
%``A Theory of Spontaneous T Violation,''
Phys.\ Rev.\ D {\bf 8} (1973) 1226.
%doi:10.1103/PhysRevD.8.1226
%%CITATION = doi:10.1103/PhysRevD.8.1226;%%
%1004 citations counted in INSPIRE as of 09 Nov 2016	


%\cite{Djouadi:2005gj}
\bibitem{Djouadi:2005gj}
A.~Djouadi,
%``The Anatomy of electro-weak symmetry breaking. II. The Higgs bosons in the minimal supersymmetric model,''
Phys.\ Rept.\  {\bf 459} (2008) 1.
%doi:10.1016/j.physrep.2007.10.005
%[hep-ph/0503173].
%%CITATION = doi:10.1016/j.physrep.2007.10.005;%%
%1021 citations counted in INSPIRE as of 09 Nov 2016	
	
%\cite{Inoue:1982pi}
\bibitem{Inoue:1982pi}
K.~Inoue, A.~Kakuto, H.~Komatsu and S.~Takeshita,
%``Aspects of Grand Unified Models with Softly Broken Supersymmetry,''
Prog.\ Theor.\ Phys.\  {\bf 68} (1982) 927;
Erratum: [Prog.\ Theor.\ Phys.\  {\bf 70} (1983) 330].
%doi:10.1143/PTP.68.927
%%CITATION = doi:10.1143/PTP.68.927;%%
%1045 citations counted in INSPIRE as of 09 Nov 2016	
	
%\cite{Langenfeld:2009tc}
\bibitem{Langenfeld:2009tc}
U.~Langenfeld, S.~Moch and P.~Uwer,
%``New results for t anti-t production at hadron colliders,''
arXiv:0907.2527 [hep-ph].
%%CITATION = ARXIV:0907.2527;%%
%122 citations counted in INSPIRE as of 28 Aug 2016
	
	
%\cite{Moch:2008qy}
\bibitem{Moch:2008qy}
S.~Moch and P.~Uwer,
%``Theoretical status and prospects for top-quark pair production at hadron colliders,''
Phys.\ Rev.\ D {\bf 78} (2008) 034003;
%5doi:10.1103/PhysRevD.78.034003
%[arXiv:0804.1476 [hep-ph]].
%%CITATION = doi:10.1103/PhysRevD.78.034003;%%
%477 citations counted in INSPIRE as of 09 Nov 2016	
%\cite{Kidonakis:2008mu}
%\bibitem{Kidonakis:2008mu}
N.~Kidonakis and R.~Vogt,
%``The Theoretical top quark cross section at the Tevatron and the LHC,''
Phys.\ Rev.\ D {\bf 78} (2008) 074005.
%doi:10.1103/PhysRevD.78.074005
%[arXiv:0805.3844 [hep-ph]].
%%CITATION = doi:10.1103/PhysRevD.78.074005;%%
%264 citations counted in INSPIRE as of 09 Nov 2016	
	
	
%\cite{Aoki:2011wd}
\bibitem{Aoki:2011wd}
M.~Aoki, R.~Guedes, S.~Kanemura, S.~Moretti, R.~Santos and K.~Yagyu,
%``Light Charged Higgs bosons at the LHC in 2HDMs,''
Phys.\ Rev.\ D {\bf 84} (2011) 055028.
%doi:10.1103/PhysRevD.84.055028
%[arXiv:1104.3178 [hep-ph]].
%%CITATION = doi:10.1103/PhysRevD.84.055028;%%
%39 citations counted in INSPIRE as of 28 Aug 2016	
	
	
%\cite{Abbiendi:2013hk}
\bibitem{Abbiendi:2013hk}
G.~Abbiendi {\it et al.} [ALEPH and DELPHI and L3 and OPAL and LEP Collaborations],
%``Search for Charged Higgs bosons: Combined Results Using LEP Data,''
Eur.\ Phys.\ J.\ C {\bf 73} (2013) 2463.
%doi:10.1140/epjc/s10052-013-2463-1
%[arXiv:1301.6065 [hep-ex]].
%%CITATION = doi:10.1140/epjc/s10052-013-2463-1;%%
%103 citations counted in INSPIRE as of 09 Nov 2016

%\cite{Chatrchyan:2012vca}
\bibitem{Chatrchyan:2012vca}
S.~Chatrchyan {\it et al.} [CMS Collaboration],
%``Search for a light charged Higgs boson in top quark decays in $pp$ collisions at $\sqrt{s}=7$ TeV,''
JHEP {\bf 1207} (2012) 143.
%doi:10.1007/JHEP07(2012)143
%[arXiv:1205.5736 [hep-ex]].
%%CITATION = doi:10.1007/JHEP07(2012)143;%%
%197 citations counted in INSPIRE as of 14 Nov 2016

%\cite{Aad:2012tj}
\bibitem{Aad:2012tj}
G.~Aad {\it et al.} [ATLAS Collaboration],
%``Search for charged Higgs bosons decaying via $H^{+} \to \tau \nu$ in top quark pair events using $pp$ collision data at $\sqrt{s}=7$ TeV with the ATLAS detector,''
JHEP {\bf 1206} (2012) 039.
%doi:10.1007/JHEP06(2012)039
%[arXiv:1204.2760 [hep-ex]].
%%CITATION = doi:10.1007/JHEP06(2012)039;%%
%209 citations counted in INSPIRE as of 14 Nov 2016


%\cite{Aad:2012rjx}
\bibitem{Aad:2012rjx}
G.~Aad {\it et al.} [ATLAS Collaboration],
%``Search for charged Higgs bosons through the violation of lepton universality in $t\bar{t}$ events using $pp$ collision data at $\sqrt{s}=7$ TeV with the ATLAS experiment,''
JHEP {\bf 1303} (2013) 076.
%doi:10.1007/JHEP03(2013)076
%[arXiv:1212.3572 [hep-ex]].
%%CITATION = doi:10.1007/JHEP03(2013)076;%%
%48 citations counted in INSPIRE as of 14 Nov 2016



%\cite{CMS:2014cdp}
\bibitem{CMS:2014cdp}
CMS Collaboration [CMS Collaboration],
%``Search for charged Higgs bosons with the H+ to tau nu decay channel in the fully hadronic final state at sqrt s = 8 TeV,''
CMS-PAS-HIG-14-020.
%%CITATION = CMS-PAS-HIG-14-020;%%
%69 citations counted in INSPIRE as of 01 Nov 2016

%\cite{TheATLAScollaboration:2013wia}
\bibitem{TheATLAScollaboration:2013wia}
The ATLAS collaboration [ATLAS Collaboration],
%``Search for charged Higgs bosons in the $\tau$+jets final state with pp collision data recorded at $\sqrt s=8$ TeV with the ATLAS experiment,''
ATLAS-CONF-2013-090.
%%CITATION = ATLAS-CONF-2013-090;%%
%90 citations counted in INSPIRE as of 03 Nov 2016v


\bibitem{kadeer}
A.~Kadeer, J.~G.~K\"orner, and M.~C.~Mauser,
%``A Phenomenological Study of Bottom Quark Fragmentation in Top Quark
%Decay,''
Eur.\ Phys.\ J.\  C {\bf 54}, 175 (2008).


\bibitem{Ali:2009sm}
A.~Ali, E.~A.~Kuraev and Y.~M.~Bystritskiy,
Eur.\ Phys.\ J.\ C {\bf 67}, 377 (2010).  %%CITATION = ARXIV:0911.3027;%%


%\cite{Li:1990cp}
\bibitem{Czarnecki}
A.~Czarnecki and S.~Davidson,
%``On the QCD corrections to the charged Higgs decay of a heavy quark,''
Phys.\ Rev.\  D {\bf 47}, 3063 (1993).
%%CITATION = PHRVA,D47,3063;%%


%\cite{Liu:1992qd}
\bibitem{Liud}
J.~Liu and Y.~P.~Yao,
%``QCD corrections to the charged Higgs boson decay of a heavy top quark,''
Phys.\ Rev.\  D {\bf 46}, 5196 (1992).
%%CITATION = PHRVA,D46,5196;%%


%\cite{Li:1990cp}
\bibitem{Li:1990cp}
C.~S.~Li and T.~C.~Yuan,
%``{QCD} Correction to Charged Higgs Decay of the Top Quark,''
Phys.\ Rev.\ D {\bf 42} (1990) 3088;
Erratum: [Phys.\ Rev.\ D {\bf 47} (1993) 2156].


%\cite{MoosaviNejad:2011yp}
\bibitem{MoosaviNejad:2011yp}
S.~M.~Moosavi Nejad,
%``B-mesons from top-quark decay in presence of the charged-Higgs boson in the Zero-Mass Variable-Flavor-Number Scheme,''
Phys.\ Rev.\ D {\bf 85} (2012) 054010;
%doi:10.1103/PhysRevD.85.054010
%[arXiv:1110.1601 [hep-ph]].
%%CITATION = doi:10.1103/PhysRevD.85.054010;%%
%5 citations counted in INSPIRE as of 28 Aug 2016
%\cite{MoosaviNejad:2012ju}
%\bibitem{MoosaviNejad:2012ju}
%S.~M.~Moosavi Nejad,
%``${\cal O}(\alpha_s)$ corrections to the B-hadron energy distribution of the top decay in the Minimal Supersymmetric Standard Model considering GM-VFN scheme,''
Eur.\ Phys.\ J.\ C {\bf 72} (2012) 2224.
%doi:10.1140/epjc/s10052-012-2224-6
%[arXiv:1205.6139 [hep-ph]].
%%CITATION = doi:10.1140/epjc/s10052-012-2224-6;%%
%6 citations counted in INSPIRE as of 14 Nov 2016


%\cite{MoosaviNejad:2016aad}
\bibitem{MoosaviNejad:2016aad}
S.~M.~Moosavi Nejad and S.~Abbaspour,
%``Next-to-leading order corrections to the hadron energy distribution from polarized top quark decay in the general two Higgs doublet model,''
arXiv:1610.03811 [hep-ph].
%%CITATION = ARXIV:1610.03811;%%

\bibitem{Cabibbo:1963yz}
N.~Cabibbo,
%``Unitary Symmetry and Leptonic Decays,''
Phys.\ Rev.\ Lett.\  {\bf 10}, 531 (1963);
%  %%CITATION = PRLTA,10,531;%%
%\cite{Kobayashi:1973fv}
%\bibitem{Kobayashi:1973fv}
M.~Kobayashi and T.~Maskawa,
%``CP Violation In The Renormalizable Theory Of Weak Interaction,''
Prog.\ Theor.\ Phys.\  {\bf 49}, 652 (1973).


%\cite{Kniehl:2012mn}
\bibitem{Kniehl:2012mn}
B.~A.~Kniehl, G.~Kramer and S.~M.~Moosavi Nejad,
%``Bottom-Flavored Hadrons from Top-Quark Decay at Next-to-Leading order in the General-Mass Variable-Flavor-Number Scheme,''
Nucl.\ Phys.\ B {\bf 862} (2012) 720.
%doi:10.1016/j.nuclphysb.2012.05.008
%[arXiv:1205.2528 [hep-ph]].
%%CITATION = doi:10.1016/j.nuclphysb.2012.05.008;%%
%12 citations counted in INSPIRE as of 28 Aug 2016


%\cite{Nejad:2013fba}
\bibitem{Nejad:2013fba}
S.~M.~Moosavi Nejad,
%``Energy spectrum of bottom- and charmed-flavored mesons from polarized top quark decay $t(?)?W^++B/D+X$ at $O(?_s)$,''
Phys.\ Rev.\ D {\bf 88} (2013) no.9,  094011.
%doi:10.1103/PhysRevD.88.094011
%[arXiv:1310.5686 [hep-ph]].
%%CITATION = doi:10.1103/PhysRevD.88.094011;%%
%7 citations counted in INSPIRE as of 25 Sep 2016



\bibitem{Nejad:2014sla}
S.~M.~Moosavi Nejad and M.~Balali,
%``Angular analysis of polarized top quark decay into $B$-mesons in two different helicity systems,''
Phys.\ Rev.\ D {\bf 90} (2014) no.11,  114017.
%Erratum: [Phys.\ Rev.\ D {\bf 93} (2016) no.11,  119904];
%doi:10.1103/PhysRevD.90.114017, 10.1103/PhysRevD.93.119904
%[arXiv:1409.1389 [hep-ph]].
%%CITATION = doi:10.1103/PhysRevD.90.114017, 10.1103/PhysRevD.93.119904;%%
%2 citations counted in INSPIRE as of 25 Sep 2016



%\cite{Nejad:2015pca}
\bibitem{Nejad:2015pca}
S.~M.~Moosavi Nejad,
%``Spin-dependent energy distribution of B-hadrons from polarized top decays considering the azimuthal correlation rate,''
Nucl.\ Phys.\ B {\bf 905} (2016) 217.
%doi:10.1016/j.nuclphysb.2016.02.014
%[arXiv:1506.08871 [hep-ph]].
%%CITATION = doi:10.1016/j.nuclphysb.2016.02.014;%%

%\cite{Nejad:2016epx}
\bibitem{Nejad:2016epx}
S.~M.~Moosavi Nejad and M.~Balali,
%``Hadron energy spectrum in polarized top quark decays considering the effects of hadron and bottom quark masses,''
Eur.\ Phys.\ J.\ C {\bf 76} (2016) no.3,  173.
%doi:10.1140/epjc/s10052-016-4017-9
%[arXiv:1602.05322 [hep-ph]].
%%CITATION = doi:10.1140/epjc/s10052-016-4017-9;%%


%\cite{Mahlon:1996pn}
\bibitem{Mahlon:1996pn} 
G.~Mahlon and S.~J.~Parke,
%``Improved spin basis for angular correlation studies in single top quark production at the Tevatron,''
Phys.\ Rev.\ D {\bf 55}, 7249 (1997).

%\cite{Kuhn:1983ix}
\bibitem{Kuhn:1983ix} 
J.~H.~K\"uhn,
Nucl.\ Phys.\ B {\bf 237}, 77 (1984);\\
J.~H.~K\"uhn, A.~Reiter and P.~M.~Zerwas,
Nucl.\ Phys.\ B {\bf 272}, 560 (1986);\\
S.~Groote and J.~G.~K\"orner,
Z.\ Phys.\ C {\bf 72} (1996) 255   [Erratum-ibid.\ C {\bf 70} (2010) 531].

\bibitem{jc}
J.~C.~Collins,~Phys. \ Rev.\ D~{\bf 66} (1998) 094002.


%\cite{Binnewies:1998vm}
\bibitem{Binnewies:1998vm} 
J.~Binnewies, B.~A.~Kniehl and G.~Kramer,
%``Inclusive $B$ meson production in $e^{+} e^{-}$ and $p \bar{p}$ collisions,''
Phys.\ Rev.\ D {\bf 58}, 034016 (1998).
%doi:10.1103/PhysRevD.58.034016
%[hep-ph/9802231].
%%CITATION = doi:10.1103/PhysRevD.58.034016;%%
%60 citations counted in INSPIRE as of 20 Nov 2016


\bibitem{dglap}
V.~N.~Gribov and L.~N.~Lipatov,
%``Deep Inelastic E P Scattering In Perturbation Theory,''
Sov.\ J.\ Nucl.\ Phys.\  {\bf 15}, 438 (1972)
[Yad.\ Fiz.\  {\bf 15}, 781 (1972)];
%%CITATION = YAFIA,15,781;%%
%\cite{Altarelli:1977zs}
%\bibitem{Altarelli:1977zs}
G.~Altarelli and G.~Parisi,
%``Asymptotic Freedom In Parton Language,''
Nucl.\ Phys.\ {\bf B126}, 298 (1977);
%%CITATION = NUPHA,B126,298;%%
%\cite{Dokshitzer:1977sg}
%\bibitem{Dokshitzer:1977sg}
Yu.~L.~Dokshitzer,
%``Calculation Of The Structure Functions For Deep Inelastic Scattering And E+
%E- Annihilation By Perturbation Theory In Quantum Chromodynamics,''
Sov.\ Phys.\ JETP {\bf 46}, 641 (1977)
[Zh.\ Eksp.\ Teor.\ Fiz.\  {\bf 73}, 1216 (1977)].
%%CITATION = ZETFA,73,1216;%%

\bibitem{Nakamura:2010zzi}
K.~Nakamura {\it et al.}\  (Particle Data Group),
%``Review of particle physics,''
J.\ Phys.\ G {\bf 37}, 075021 (2010).
\bibitem{Kniehl:2008zza}
B.~A.~Kniehl, G.~Kramer, I.~Schienbein, and H.~Spiesberger,
%``Finite-mass effects on inclusive $B$ meson hadroproduction,''
Phys.\ Rev.\  D {\bf 77}, 014011 (2008).
%  [arXiv:0705.4392 [hep-ph]].
%%CITATION = PHRVA,D77,014011;%%

%\cite{Abbiendi:2002vt}
\bibitem{Abbiendi:2002vt}
G.~Abbiendi {\it et al.}\  (OPAL Collaboration),
%``Inclusive analysis of the b quark fragmentation function in Z decays at
%LEP. ((B)),''
Eur.\ Phys.\ J.\  C {\bf 29}, 463 (2003).
%  [arXiv:hep-ex/0210031].
%%CITATION = EPHJA,C29,463;%%


\bibitem{Heister:2001jg}
A.~Heister {\it et al.}\  (ALEPH Collaboration),
%``Study of the fragmentation of b quarks into B mesons at the Z peak,''
Phys.\ Lett.\  B {\bf 512}, 30 (2001).
%  [arXiv:hep-ex/0106051].
%%CITATION = PHLTA,B512,30;%%

%\cite{Abe:1999ki}
\bibitem{Abe:1999ki}
K.~Abe {\it et al.}\  (SLD Collaboration),
%``Precise measurement of the b-quark fragmentation function in Z0 boson
%decays,''
Phys.\ Rev.\ Lett.\  {\bf 84}, 4300 (2000);
%  [arXiv:hep-ex/9912058].
%%CITATION = PRLTA,84,4300;%%
%\cite{Abe:2002iq}
%\bibitem{Abe:2002iq}
%  K.~Abe {\it et al.}  [SLD Collaboration],
%``Measurement of the b-quark fragmentation function in Z0 decays,''
Phys.\ Rev.\  D {\bf 65}, 092006 (2002);
%  [Erratum-ibid.\  D
{\bf 66}, 079905 (2002).
%]
%  [arXiv:hep-ex/0202031].
%%CITATION = PHRVA,D65,092006;%%

%\cite{Soleymaninia:2013cxa}
\bibitem{Soleymaninia:2013cxa}
M.~Soleymaninia, A.~N.~Khorramian, S.~M.~Moosavi Nejad and F.~Arbabifar,
%``Determination of pion and kaon fragmentation functions including spin asymmetries data in a global analysis,''
Phys.\ Rev.\ D {\bf 88} (2013) no.5,  054019.
%Addendum: [Phys.\ Rev.\ D {\bf 89} (2014) no.3,  039901]
%doi:10.1103/PhysRevD.88.054019, 10.1103/PhysRevD.89.039901
%[arXiv:1306.1612 [hep-ph]].
%%CITATION = doi:10.1103/PhysRevD.88.054019, 10.1103/PhysRevD.89.039901;%%
%17 citations counted in INSPIRE as of 28 Sep 2016

%\cite{Nejad:2015fdh}
\bibitem{Nejad:2015fdh}
S.~M.~Moosavi Nejad, M.~Soleymaninia and A.~Maktoubian,
%``Proton fragmentation functions considering finite-mass corrections,''
Eur.\ Phys.\ J.\ A {\bf 52} (2016) no.10,  316.
%doi:10.1140/epja/i2016-16316-6
%[arXiv:1512.01855 [hep-ph]].
%%CITATION = doi:10.1140/epja/i2016-16316-6;%%
%3 citations counted in INSPIRE as of 19 Nov 2016

	

\bibitem{collins}
J.~C.~Collins,
%``Finite-mass effects on inclusive $B$ meson hadroproduction,''
Phys.\ Rev.\  D {\bf 58}, 094002 (1998).

\end{thebibliography}
\end{document}